\documentclass[a4paper,11pt]{article}
\usepackage[top=4cm, bottom=2.7cm, left=2.5cm, right=2.5cm, headsep=60pt]{geometry}
\usepackage{amsmath, amssymb, amsthm}
\usepackage{graphicx}
\usepackage{caption}
\usepackage{times}
\usepackage{titlesec}
\usepackage{fancyhdr}

\titleformat{\section}
  {\normalfont\Large\bfseries} 
  {\thesection.}{1em}{} 
\titlespacing*{\section}
  {0pt}{1\baselineskip}{0pt} 

\titleformat{\subsection}
  {\normalfont\large\itshape} 
  {\thesubsection.}{1em}{} 
\titlespacing*{\subsection}
  {0pt}{1\baselineskip}{0pt} 

\titleformat{\subsubsection}[runin]
  {\normalfont\normalsize\itshape} 
  {\thesubsubsection.}{1em}{} 
\titlespacing*{\subsubsection}
  {0pt}{0pt}{0.5em} 

\begin{document}

\vspace{1em}

\begin{center}
  { \fontsize{16}{10} \textbf{Reduced Basis Model for Compressible Flow} } \\
    \vspace{0.5cm}
    \normalsize \textbf{Marian Staggl$^{1}$, Wolfgang Sanz$^{1}$, Paul Pieringer$^{2}$} \\ 
    \vspace{0.2cm}
  {\fontsize{9}{10}\selectfont
  \textit{$^1$ Institute of Thermal Turbomachinery and Machine Dynamics, Graz University of Technology, Austria}\\
  \textit{$^2$ ABES Pircher \& Partner GmbH, Graz, Austria}\\} 
  { \fontsize{9}{10}\textit{Corresponding author: Marian Staggl (marian.staggl@tugraz.at)}} \\
\end{center}

\vspace{1em}


\vspace{1em}

\noindent\textbf{Abstract.} Numerical simulations are a valuable research and layout tool for fluid flow problems, yet repeated evaluations of parametrized problems, necessary to solve optimization problems, can be very costly. One option to speed up this process is to replace the costly CFD model with a cheaper one. These surrogate models can be either data-driven or they can also rely on reduced basis (RB) methods to speed up the calculations. In contrast to data-driven surrogate models, the latter are not based on regression techniques but are still aimed at explicitly solving the conservation equations. Their speed-up comes from a strong reduction of the solution space, which results in much smaller algebraic systems that need to be solved. Within this work, an RB model, suited for slightly compressible flow, is presented and tested on different flow configurations. The model is stabilized using a Petrov-Galerkin method with trial and test function spaces of different dimensionality to generate stable results for a wide range of Reynolds numbers. The presented model applies to geometrically and physically parametrized flow problems. Finally, a data-driven approach was used to extend it to turbulent flows.

\vspace{1em}
\section{Introduction}
\label{sec:Introduction}
\subsection{RB methodology}
While data-driven surrogate models use correlations to predict outcomes for unseen parametric combinations, RB models enable the speed up of numerical calculations by reducing the dimensionality of the solution space while still solving the governing equations. Starting from a flow problem which is dependent on a set of parameters $\bar{\mu}$, RB models assume that there exists a small set of coherent flow structures that, if appropriately superimposed, are sufficient to reproduce any result of this parametrized problem with high accuracy. This assumption can be expressed by Eq.~\ref{eq:continuous_solution_space} where the coherent flow structures are given by $\varphi$ and are independent of parameter values. 

\begin{equation}
	\label{eq:continuous_solution_space}
	s(x, \mu) = \sum_{i=1}^{n} \varphi_{i}(x) \, r_{i}(\mu)
\end{equation}

Throughout this work, these flow structures will be referred to as modes. The goal of an RB model is thus to find the weighting factors $r(\mu)$ of the different flow structures such that their superimposition fulfils the governing equations as well as possible. A linear PDE system is assumed to outline the basic principle in the first place. Discretizing the PDE system gives rise to a large algebraic system of size $h \times h$ where $h$ is given by the number of nodes times the number of variables (see Eq.~\ref{eq:high_fidelity_system}).

\begin{equation}
    \label{eq:high_fidelity_system}
    \bar{\bar{N}}(\mu) \, \bar{s}(\mu) = \bar{f}(\mu)
\end{equation}

In its discrete form, the solution and the modes may be expressed as a vector $\bar{s}$, with each entry being a flow variable on a node (or at the cell centre) of a chosen computational mesh. In this case, Eq.~\ref{eq:continuous_solution_space} has a discrete representation shown in Eq.~\ref{eq:discrete_solution_space}

\begin{equation}
	\label{eq:discrete_solution_space}
	\bar{s}(\mu) = \bar{\varphi}^{i} \, r^{i}(\mu) = \bar{\bar{\Phi}} \, \bar{r}
\end{equation}

If the identity of Eq.~\ref{eq:discrete_solution_space} is introduced into the algebraic system of Eq.~\ref{eq:high_fidelity_system}, the number of DoFs of the initial system reduces to the number of modes (see Eq.~\ref{eq:reduced_system_1}). This algebraic system now has $h$ equations but only $n$ DoFs.

\begin{equation}
	\label{eq:reduced_system_1}
	\bar{\bar{N}} \, \bar{\bar{\Phi}} \, \bar{r} = \bar{f}
\end{equation}

A Galerkin projection is used to reduce the number of equations. The test function space is chosen to be equal to the trial function space, and thus, the system is tested against the set of modes (see Eq.~\ref{eq:reduced_system_2}). As the modes are independent of the parameter values, some matrix multiplications can be carried out in advance, leading to a $n \times n$ system that needs to be solved.

\begin{equation}
	\label{eq:reduced_system_2}
	\bar{\bar{\Phi}}^{T}\bar{\bar{N}} \, \bar{\bar{\Phi}} \, \bar{r} = \bar{\bar{\Phi}}^{T} \, \bar{f} \quad \rightarrow \quad \bar{\bar{R}} \, \bar{r} = \bar{g}
\end{equation}

The resulting algebraic system is not dependent of the number of mesh nodes but on the number of modes. This means that if a small number of modes can be found suitable to reconstruct solutions with high accuracy, the system of Eq.~\ref{eq:reduced_system_2} can provide fast and accurate predictions for a specific parametrized problem. What is important, though, is to effectively decouple the time-consuming steps during the setup of the model (offline phase) from the cheap evaluation of the model during use (online phase).
\\An efficient derivation of these modes is thus an essential building block for the setup of an RB model. While there are several options to do so, the most widespread one is the proper orthogonal decomposition (POD) \cite{Hijazi.2020, Ali.2018, Quarteroni.2016, Ballarin.2015, Wang.2012,Bergmann.2009, Rowley.2004, Carlberg.2013}. The method relies on a singular value decomposition of available flow results and represents a fast and efficient way to derive an optimal set of modes.
\\The reduced RB methodology applies to a wide range of problems, including linear partial differential equations (PDE), Stokes flow \cite{Ali.2018} and also non-linear systems like the Navier-Stokes equations. Concerning fluid flow examples, the treatment of the incompressible Navier-Stokes equations was investigated by many authors \cite{Bergmann.2009, Ballarin.2015, Quarteroni.2016, Ali.2018, Akkari.2019, Hijazi.2020, Nakamura.2024} but also the solution of compressible flow problems were investigated with promising results. Rowley et al. \cite{Rowley.2001, Rowley.2004} introduced a simplified version of the compressible Navier-Stokes equation by neglecting the energy equation and closing the system with isentropic relations. Other formulations of RB models for compressible flow can be found, e.g. in \cite{Tezaur.2016}. Concerning the simulation of turbulent flows, an incompressible RB model was presented by \cite{Hijazi.2020}, which relies on a data-driven approach to model the viscosity fields and uses a finite volume approach to build the algebraic system. Overall, methods to derive the high-fidelity system (see Eq.~\ref{eq:high_fidelity_system}) rely mainly on finite element (FE) or finite volume (FV) approaches. 
\\One drawback of RB methods is that they can suffer from stability issues. There are two primary sources of instabilities where one concerns especially the incompressible formulation of the flow equations. In the case of incompressibility, the velocity fields are divergence-free, and as linear operations are used to construct the modal base for RB models, the velocity field of these modes is divergence-free as well and fulfils the continuity equation automatically. Due to this, it is no longer possible to recover the pressure fields accurately. In practice, calculating the pressure field without further stabilization will result in spurious pressure oscillations \cite{Ali.2018} and may cause the model to converge to a wrong solution \cite{Ballarin.2015}. Several different stabilization methods have been proposed, depending on the discretization approach. One way to stabilize RB models, derived by a finite element discretization, is to enrich the velocity solution space with non-divergence-free modes \cite{Ballarin.2015, Ali.2018, Hijazi.2020} (so called supremizer solutions). Following methodologies from the FE framework, it is also possible to stabilize the RB model using penalty methods like least squares Petrov Galerkin (LSPG) or streamwise upwind Petrov Galerkin (SUPG). In both cases, an additional term is added to the Galerkin contribution, changing the test space slightly \cite{Bergmann.2009, Quarteroni.2016, Ali.2018}. Alternatively, in the finite volume method, the system of equations may also be closed using a pressure Poisson equation \cite{Stabile.2018}.
\\ Convection-dominated flows at high Reynolds numbers introduce another source of instabilities. A reason why especially RB models suffer from them is due to a reduced dissipation caused by the truncated higher order modes \cite{Bergmann.2009, COUPLET.2003, Wang.2012}. This leads to a closure problem similar to the one in Large Eddy Simulations (LES), where small dissipative scales are unresolved. Even though the contribution of small-scale modes to the accurate reconstruction of flow results is negligible, they significantly influence energy dissipation. The lack of dissipative modes can lead to unstable long-term behaviour \cite{Bergmann.2009, Wang.2012, COUPLET.2003} if their effects are not modelled in another way. Again, there are multiple options to mitigate the effects of missing dissipation. The viscosity can be increased artificially to account for the missing dissipative modes. This can be done by increasing the overall viscosity or by linking the viscosity to the mode number \cite{Wang.2012} or by using eddy viscosity fields when setting up the RB model analogue to the procedure in Reynolds Averaged Navier Stokes equations (RANS) \cite{Hijazi.2020}. Another possibility is to use stabilization procedures like LSPG, stemming from the finite element framework. As shown by \cite{Bergmann.2009, Carlberg.2013, Tezaur.2016}, both methods can help to stabilize the RB model. Finally, it is possible to modify the velocity modes such that they introduce more dissipation \cite{Bergmann.2009, Akkari.2019}.

\section{Methodology}
\subsection{Isentropic Navier Stokes Equations}
The model presented in this work is based on a simplified version of the compressible Navier-Stokes equations. The equations were initially proposed by Rowley et al. \cite{Rowley.2001} and are valid for moderate Mach numbers and small temperature gradients. Rowley et al. \cite{Rowley.2001} derived them by adding a viscous term to the isentropic Euler equations. Still, they may also be derived from the compressible Navier Stokes equations, as shown below. In their compressible form, the stationary momentum and continuity equations without volumetric forces read:

\begin{equation}
    \label{eq:compressible_momentum}
    u_{j}\frac {\partial u_{i}}{\partial x_{j}}+{\frac{1}{\rho } \frac {\partial p}{\partial x_{i}}}-{\frac{1}{\rho } \frac {\partial }{\partial x_{j}}}\left[\mu \left({\frac {\partial u_{i}}{\partial x_{j}}}+{\frac {\partial u_{j}}{\partial x_{i}}}-{\frac {2}{3}}\delta _{ij}{\frac {\partial u_{l}}{\partial x_{l}}}\right)\right]=0
\end{equation}
\begin{equation}
    \label{eq:compressible_continuity}
    \frac{\partial}{\partial x_{i}}\left(\rho u_{i}\right)=0
\end{equation}

To this point, the system given by Eq.~\ref{eq:compressible_momentum} and Eq.~\ref{eq:compressible_continuity} is not closed as the energy equation is missing. To close the system, the isentropic relations given in Eq.~\ref{eq:isentropic_relations} are used, where $p$ and $\rho$ are pressure and density, and $a$ is the local speed of sound.

\begin{equation}
    \label{eq:isentropic_relations}
    \begin{aligned}
        \frac{\rho}{\rho_{1}}=\left(\frac{a}{a_{1}}\right)^{\frac{2}{\kappa-1}} \quad \frac{p}{p_{1}}=\left(\frac{a}{a_{1}}\right)^{\frac{2\kappa}{\kappa-1}}
    \end{aligned}
\end{equation}

If the isentropic relations from Eq.~\ref{eq:isentropic_relations} are inserted into the continuity Eq.~\ref{eq:compressible_continuity}, one obtains Eq.~\ref{eq:compressible_continuity}. With regard to the momentum equation, it is further assumed that even though the fluid is compressible, the divergence of the velocity field is small. Following this assumption and combining Eq.~\ref{eq:compressible_momentum} with Eq.~\ref{eq:isentropic_relations}, this results in a new form of the momentum equation given in Eq.~\ref{eq:isentropic_momentum1}.

\begin{equation}
    \label{eq:isentropic_momentum1}
    u_{j} \frac{\partial u_{i}}{\partial x_{j}} + \frac{2}{\kappa - 1} a \frac{\partial a}{\partial x_{i}} - {\frac{1}{\rho } \frac {\partial }{\partial x_{j}}}\left[\mu \left({\frac {\partial u_{i}}{\partial x_{j}}}+{\frac {\partial u_{j}}{\partial x_{i}}} \right)\right] = 0
\end{equation}
\begin{equation}
    \label{eq:isentropic_continuity}
    u_{j} \frac{\partial a}{\partial x_{j}} + \frac{\kappa - 1}{2} a \frac{\partial u_{j}}{\partial x_{j}} = 0\\
\end{equation}

By applying the product rule to the viscous terms and again assuming that the divergence of the velocity field is close to zero, Eq.~\ref{eq:isentropic_momentum1} can be simplified further. Finally, for now, it is assumed that the viscosity is constant over the domain, which leads to Eq.~\ref{eq:isentropic_momentum2}.

\begin{equation}
    \label{eq:isentropic_momentum2}
    u_{j} \frac{\partial u_{i}}{\partial x_{j}} + \frac{2}{\kappa - 1} a \frac{\partial a}{\partial x_{i}} - \nu \frac{\partial^{2} u_{i}}{\partial x_{j}^{2}} = 0
\end{equation}

This system of equations was chosen for two main reasons. First, it allows the expression of all flow variables in the same physical dimension, which is essential for the POD, as will be shown later. Furthermore, in contrast to the fully compressible Navier-Stokes equations, it only has a non-linearity of the second degree, facilitating the implementation of a fast and efficient RB model.

\subsection{Introducing the Example Case}
\begin{figure}[!htpb]
	\centering
	\includegraphics[width=\textwidth]{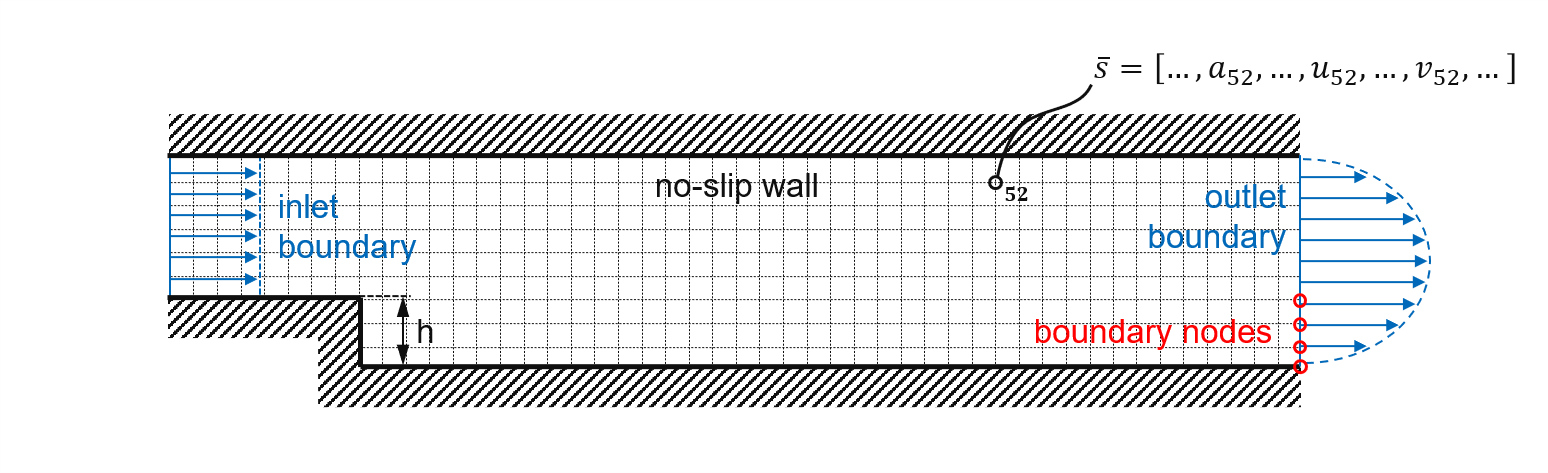}
	\caption{Sudden expansion channel as RB model example case.}
	\label{fig:sketch_sudden_expansion_channel}
\end{figure}
RB models are often applied to solve canonical problems, e.g., a Karman vortex street \cite{Hijazi.2020, Wang.2012} for in-stationary flows or a sudden expansion channel for stationary flows \cite{Ali.2018, Quarteroni.2016, Ballarin.2015, Bergmann.2009}. To enhance the comparability of the results of this work with results from the literature, the sudden expansion channel is chosen as an exemplary case to validate the implemented RB method. A treatment of this case for a laminar incompressible flow can be found in \cite{Quarteroni.2016, Ballarin.2015} and the same case for an incompressible turbulent flow was investigated by Hijazi et al. \cite{Hijazi.2020}. Furthermore, applied methods can be illustrated well using this example case. A sketch of the flow problem is shown in Figure \ref{fig:sketch_sudden_expansion_channel}. The flow setup is defined by an inlet velocity boundary to the left and a pressure outlet boundary condition to the right, while the walls are defined as a no-slip boundary condition. To enforce the boundary conditions, velocity values are fixed at the inlet boundary nodes, while the local speed of sound is fixed at the outlet boundary nodes. The flow separates at the corner and reattaches further downstream before reaching the channel outlet. The inlet velocity and the height of the step determine the exact reattachment position. 
\\The problem is thus defined in a two-dimensional parameter space where a variable inlet velocity will be referred to as a physical parameter. At the same time, the step height represents a geometrical parameter. This distinction is important as physical and geometrical parameters are treated significantly differently, as will be shown. Each solution of this parametrized problem can be expressed in terms of a vector $\bar{s}(\mu)$ where the ID of each node of the computational mesh specifies the index position within this solution vector.
A MATLAB FE code is used to carry out 400 CFD simulations of this sudden expansion channel using different geometries and boundary conditions, using a random approach to sample the parameter space. This train set serves as a basis for determining the solution space for the RB model via proper orthogonal decomposition. A further test set of 200 calculations is created to evaluate the model's accuracy.

\subsection{Proper Orthogonal Decomposition}
A crucial step in creating an efficient and accurate RB model is the identification of the modes $\varphi(x)$, respectively, their discrete counterparts $\bar{\varphi}$ shown in Eq.~\ref{eq:discrete_solution_space}. The goal is to choose these modes such that a small number is sufficient to accurately reconstruct available solutions $\bar{s}(\mu)$ of a parametrized flow problem. Even though multiple methods exist to do so, the proper orthogonal decomposition (POD) is by far the most popular one \cite{Hijazi.2020, Ali.2018, Quarteroni.2016, Ballarin.2015, Wang.2012, Bergmann.2009} and is used throughout this work. 
\\If the investigated solution $\bar{s}$ stems from a numerical simulation, it can be represented in the discrete form of a vector (see Eq.~\ref{eq:discrete_solution}). Each vector entry is associated with a specific node (of course, this is only true for FE methods) of the computational mesh and a specific flow variable (e.g. x-velocity, y-velocity, local speed of sound, ...).

\begin{equation}
    \label{eq:discrete_solution}
    \bar{s} = [\bar{a}; \, \bar{u}; \, \bar{v}]
\end{equation}

To derive the POD modes, a set of $m$ solutions of a parametrized flow problem are necessary (for the given example, 400 are available). The solutions to this parametrized problem must be available on the same numerical mesh (same topology) so that the vector representations of the different solutions are equivalent; however, mesh deformations are possible. As a next step, the solutions are collected in a snapshot matrix (see Eq.~\ref{eq:snapshot_matrix}).

\begin{equation}
    \label{eq:snapshot_matrix}
    \bar{\bar{S}}_{h \times m} = [\bar{s}_{1}, \bar{s}_{2}, \bar{s}_{3}, ..., \bar{s}_{m}]
\end{equation}

By squaring this snapshot matrix, one obtains a covariance matrix (see Eq.~\ref{eq:covariance_matrix}), where each entry of this covariance matrix contains a measure for the correlation between two solutions in terms of a scalar product being a discrete version of an inner product.

\begin{equation}
    \label{eq:covariance_matrix}
   \bar{\bar{C}} = \bar{\bar{S}} \cdot \bar{\bar{S}}^{\intercal} 
\end{equation}

As a next step, the eigenvalues and eigenvectors of the covariance matrix are calculated and sorted according to their eigenvalue $\lambda$; the variable $\bar{\psi}$ denotes an eigenvector. Finally, the identified eigenvectors are projected onto the snapshot matrix (see Eq.~\ref{eq:mode_calculation}), resulting in the $i^{th}$ POD mode.

\begin{equation}
    \label{eq:eigenvalue_calculation}
   \bar{\bar{C}} \cdot \bar{\psi} = \lambda \cdot \bar{\psi} 
\end{equation}
\begin{equation}
    \label{eq:mode_calculation}
   \bar{\varphi}_i = \frac{1}{\sqrt{\lambda}_i} \, \bar{\bar{S}} \cdot \bar{\psi}_i
\end{equation}

\begin{figure}[!htbp]
	\centering
	\includegraphics[width=\textwidth]{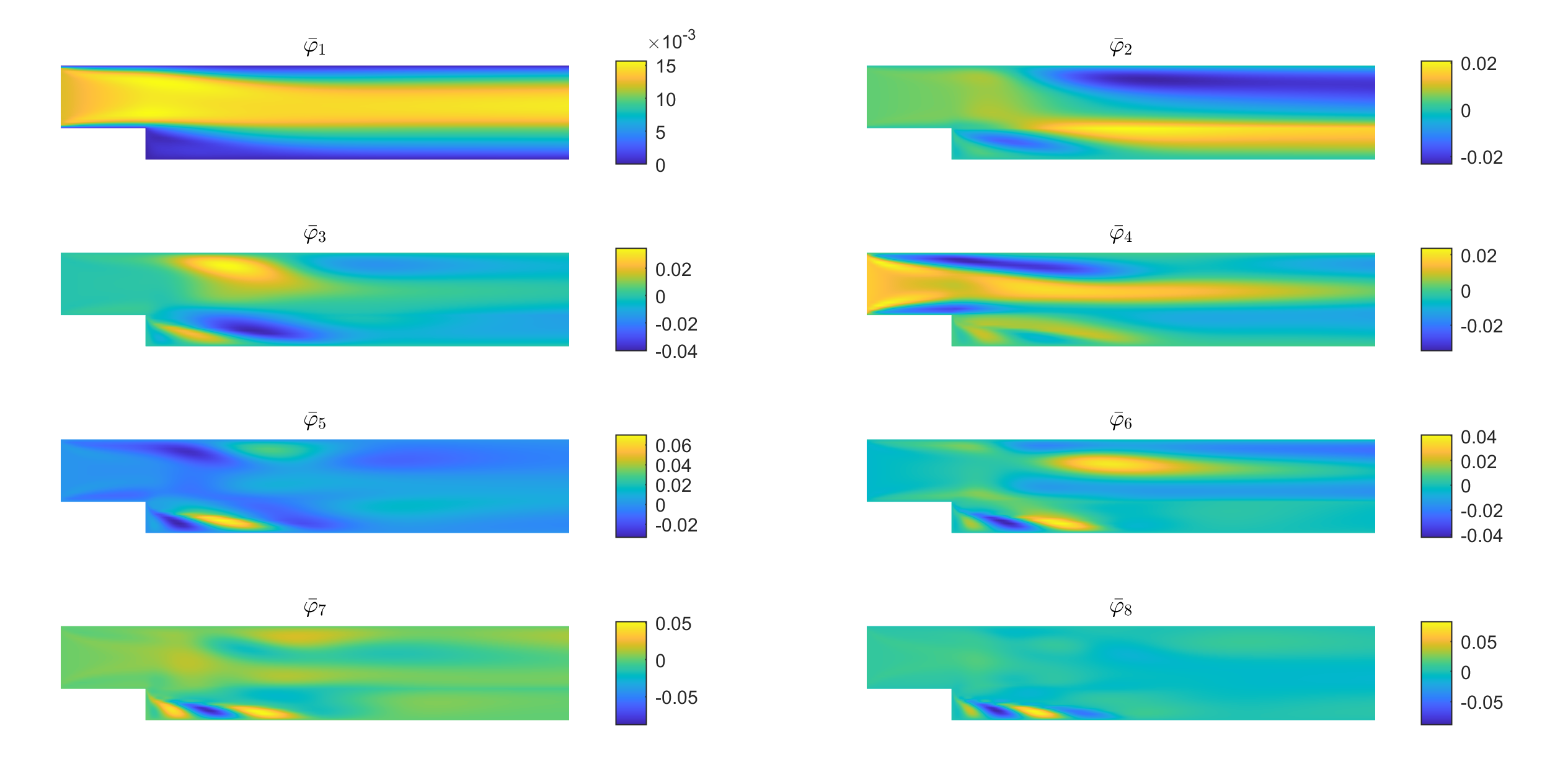}
	\caption{Velocity magnitude POD modes \(\varphi_{1}\) to \(\varphi_{8}\) for the presented sudden expansion channel. Higher modes tend to contain smaller structures.}
	\label{fig:vel_POD_modes}
\end{figure}

The POD modes are collected in a matrix. Usually, not all $m$ modes are considered but only a subset $n$ which leads to the required RB base $\bar{\bar{\Phi}}=[\bar{\varphi}_1, \, \bar{\varphi}_2, \, ..., \,\bar{\varphi}_n]$ where each eigenvector has $h$ entries. 
\\To give an illustrative example, Figure \ref{fig:vel_POD_modes} shows the first eight velocity POD modes of the example case. In this case, only the velocity fields were subject to the decomposition, not the full snapshot matrix. In this case, the entries of the covariance matrix have the physical dimension of a velocity-squared and thus are proportional to the kinetic energy of the fluid (if only small density changes are assumed) connecting the POD to a physical measure.
\\However, if the full snapshot matrix is subject to the POD, this leads to a problem with regard to physical dimensions. A flow field is not defined by a single variable but by multiple variables (see Eq.~\ref{eq:discrete_solution}), including the velocity components $u, v$ and e.g. in the incompressible case $p$. Recalling POD methodology means that the entries of the covariance matrix would contain mixed dimensions (see Eq.~\ref{eq:standard_scalar_prod}). This makes the decomposition somewhat arbitrary because variables with high magnitude are weighted disproportionately.

\begin{equation}
    \label{eq:standard_scalar_prod}
    C_{(1, 2)} = \bar{s}_{1} \cdot \bar{s}_{2} = \bar{p}_{1} \cdot \bar{p}_{2} + \bar{u}_{1} \cdot \bar{u}_{2} + \bar{v}_{1} \cdot \bar{v}_{2}
\end{equation}

 One way to circumvent this problem is to decompose pressure and velocity fields separately \cite{Quarteroni.2016, Ballarin.2015, Ali.2018} but this leads to stability problems even in the case of low Reynolds numbers and requires further stabilization. Another possibility is to use compatible physical dimensions like $u, v$ and $a$, which is possible for the chosen system of equations. This leads to a more consistent meaning of the entries in the covariance matrix (see Eq.~\ref{eq:dimentionally_consistent_inner_prod}).

 \begin{equation}
    \label{eq:dimentionally_consistent_inner_prod}
    C_{(1, 2)} = \bar{s}_{1} \cdot \bar{s}_{2} = \bar{a}_{1} \cdot \bar{a}_{2} + \bar{u}_{1} \cdot \bar{u}_{2} + \bar{v}_{1} \cdot \bar{v}_{2}
\end{equation}

Furthermore, it is possible to weigh some parts of this scalar product and thus transform it in such a way that each entry of the covariance matrix contains the integral of the stagnation enthalpy (see Eq.~\ref{eq:energy_based_prod}). This procedure was proposed by Rowley et al. \cite{Rowley.2004}, and it allows linking the POD with physical values.

\begin{equation}
    \label{eq:energy_based_prod}
   C_{(1, 2)} \approx h_{stag} =\int (u_{1} \cdot u_{2} + v_{1} \cdot v_{2} + \frac{1}{\kappa - 1} a_{1} \cdot a_{2}) \cdot dV
\end{equation}

\subsection{Boundary Treatment}
There are two main approaches to implement boundary conditions for RB models. The first approach is based on a penalty method where additional residual terms are introduced, which take large values if the desired boundary conditions are not fulfilled. This penalty term is weighted with a factor, and the higher its value, the more precisely the boundary conditions are met \cite{Hijazi.2020}. 
\\Another approach is a so-called lifting function, which is used to homogenize derived POD modes at essential boundary conditions \cite{Gunzburger.2007, Hijazi.2020}. After this step, the RB base functions are orthogonal to nodes of the essential boundary conditions, and their strength may be altered without changing values at the boundaries. A simple illustrative example of homogenized boundary conditions is given by the case of flow solutions stemming from a non-stationary flow with constant boundary conditions. Here, the boundary conditions are homogenized by removing the mean flow solution. By doing so, all derived POD modes are orthogonal to the essential boundary conditions \cite{Bergmann.2009}, but the derived mean mode needs to be added to the base functions. While the RB model alters the amplitudes of the homogenized POD modes, the mean-mode strength is kept at a constant value of one. For more complex cases, lifting functions are derived in other ways, e.g. shown by Hijazi et al. \cite{Hijazi.2020}. There, lifting functions are introduced, which are homogeneous everywhere except in one entry of the POD mode, where they are set to a unitary value. This is equivalent to setting the results to zero at the boundary nodes and adding the unitary vectors to the POD base. Finally, Gunzburger et al. \cite{Gunzburger.2007} proposed a QR decomposition to homogenize boundary conditions. In this work, an approach similar to \cite{Gunzburger.2007} was chosen, but it is based on a singular value decomposition rather than a QR decomposition. 
\\Eq.~\ref{eq:boundary_modes} shows how a solution to the parametrized problem can be expressed as a linear superposition of POD modes represented by the base $\bar{\bar{\Phi}}$. To impose boundary conditions, the initial POD modes are split up into two sets of modes where one set $\bar{\bar{\Omega}}$ is non-homogeneous at the essential boundaries while the second set $\bar{\bar{\Psi}}$ is. This means that the entries of the vector $\bar{b}$ can be used to set the boundary conditions. In contrast, the entries $\bar{d}$ can be changed without altering values at the essential boundaries and will be determined by solving the governing equations. 

\begin{equation}
    \label{eq:boundary_modes}
    \bar{s}(\mu) = \bar{\bar{\Phi}} \, \bar{r} = \bar{\bar{\Omega}} \, \bar{b} + \bar{\bar{\Psi}} \, \bar{d}
\end{equation}

The decomposition is carried out in two steps to create the two sets of modes. In the first step, only vector entries associated with essential boundary conditions are considered (see Eq.~\ref{eq:zonal_svd}). This is expressed by the operation $\bar{\bar{P}}^{T} \, \bar{\bar{S}}$ where the columns of the matrix $\bar{\bar{P}}$ are a subset of the identity matrix. In this way, only specific rows of the snapshot matrix are selected (e.g. the tenth column of the identity matrix selects only the tenth row of the snapshot matrix). An SVD of this product is carried out to derive the right eigenvectors (see Eq.~\ref{eq:zonal_svd}).

\begin{equation}
    \label{eq:zonal_svd}
    \_, \_, \bar{\bar{E}} = svd\left(\bar{\bar{P}}^{T} \, \bar{\bar{S}}\right)
\end{equation}

The right eigenvectors are projected onto the initial snapshot matrix which results in a set of modes that exhibit a fast decay at the essential boundary conditions. At the same time, their amplitude remains non-zero within the domain (unless the results stem from a linear system of equations).

\begin{equation}
    \label{eq:split_snapshot_matrix}
    \bar{\bar{\Omega}}=\bar{\bar{S}} \, \bar{\bar{E}}
\end{equation}

The first $nb$ columns of $\bar{\bar{\Omega}}$ are then chosen as boundary modes. In the case of linearly decomposable boundary conditions, the singular values of the SVD eventually reach zero after a finite number of modes. In this case, the outlined procedure does indeed generate two sets of modes where one has exactly homogeneous boundary conditions, and it also yields a clear choice for $nb$. However, if the boundary conditions are not linearly decomposable, the right choice of $nb$ depends on the investigated problem. In the second step, the remaining $n-nb$ modes are decomposed using the energy-based inner product, defined in Eq.~\ref{eq:energy_based_prod} which gives the second set of modes $\bar{\bar{\Psi}}$.
\\If Eq.~\ref{eq:boundary_modes} is multiplied with the sampling matrix $\bar{\bar{P}}$ it changes to Eq.~\ref{eq:imposing_boundary_conditions1} where due to the orthogonality of the domain modes, one term drops out of the equation.

\begin{equation}
    \label{eq:imposing_boundary_conditions1}
    \bar{\bar{P}}^{T} \bar{s}(\mu) = \bar{\bar{P}}^{T}\bar{\bar{\Omega}}\bar{b} + \bar{\bar{P}}^{T}\bar{\bar{\Psi}}\bar{d} \quad \rightarrow \quad \bar{\bar{P}}^{T}\bar{\bar{\Psi}}\bar{d} = 0
\end{equation}

If the RB model should be evaluated for new parameter values, the values of the flow variables at the essential boundaries are known, and thus the product $\bar{\bar{P}}^{T} \bar{s}(\mu)$ is known, even if the complete solution $\bar{s}(\mu)$ is unknown. The boundary conditions for this new parameter set are imposed via a least squares approach: Eq.~\ref{eq:imposing_boundary_conditions1} is rearranged to Eq.~\ref{eq:imposing_boundary_conditions2} where $+$ implies the Moore-Penrose pseudo inverse \cite{Penrose.1955}, finally yielding the amplitudes of the boundary modes. In literature, this approach is also known as Gappy POD.

\begin{equation}
    \label{eq:imposing_boundary_conditions2}
    \bar{b} = (\bar{\bar{P}} \, \bar{\bar{\Omega}})^{+} \, \bar{\bar{P}} \, \bar{s}
\end{equation}

\subsection{RB system assembly}
\begin{figure}[!htpb]
    \vspace{0.8cm}
	\centering
	\includegraphics[width=\textwidth]{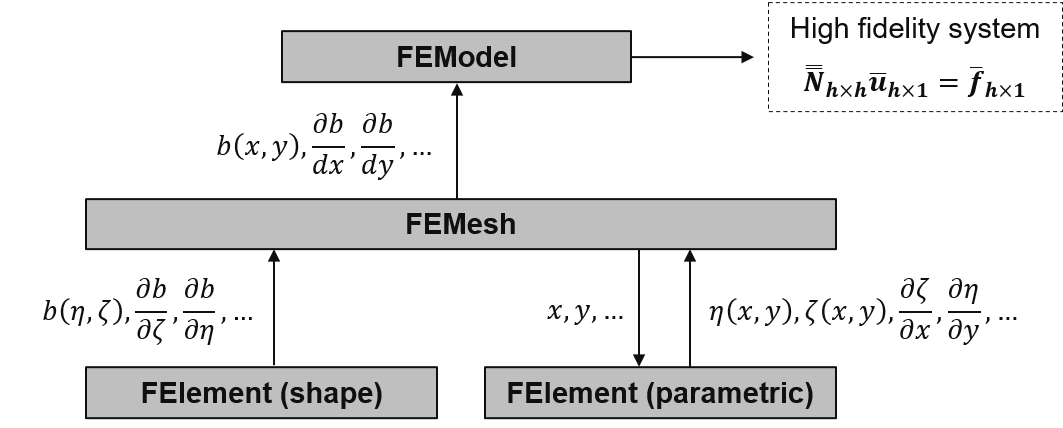}
	\caption{Structure of implemented finite element code with base functions $b$ defined on reference coordinates $\eta, \zeta$ and global coordinates $x, y$.}
	\label{fig:matlab_fe_structure}
\end{figure}

The remaining unknowns $\bar{d}$ must be determined by solving the governing equations in reduced form (see Eq.~\ref{eq:reduced_system_2}). An essential part of setting up the equations is assembling the high-fidelity system expressed by $\bar{\bar{N}}$. For this purpose, an FE solver was implemented in MATLAB, and the basic structure of the code is shown in Figure \ref{fig:matlab_fe_structure}. The solver's primary class is the FElements, which provide base functions and integration weights; a FEMesh, which maps the elements and calculates the global base function derivatives; and finally, an FEModel, where the system matrices for the isentropic Navier Stokes equations are assembled (see top right).
The system of isentropic Navier-Stokes equations is given by the continuity equation shown in Eq.~\ref{eq:isentropic_continuity}, and the momentum equation shown in Eq.~\ref{eq:isentropic_momentum2}. Parts of the equations are due to linear terms, while other parts of the equations are given by non-linear terms with a polynomial non-linearity of the second degree. In Eq.~\ref{eq:isentropic_navier_stokes_ql}, the linear terms are expressed by the operator $L$, and the non-linear terms are given by the operator $Q$.

\begin{equation}
    \label{eq:isentropic_navier_stokes_ql}
    Q(\bar{s}, \bar{s}) - L(\bar{s}) = 0
\end{equation}

The linear terms arise from the viscous forces and are given in Eq.~\ref{eq:linear_terms}, where the zero entry corresponds to the continuity equation while the second term is due to the momentum equations. The non-linear terms are given in Eq.~\ref{eq:quadratic_terms}, and again, the first entry arises from the continuity equation while the second comes from the momentum equation.

\vspace{0.2cm}
\begin{equation}
    \label{eq:linear_terms}
    L(\bar{s}) = \left( 0, \quad
    \nu \frac{\partial^{2} u_{i}}{\partial x_{j}^{2}} \right)
\end{equation}
\vspace{0.2cm}
\begin{equation}
    \label{eq:quadratic_terms}
    Q(\bar{s},\bar{s}) = \left(u_{j} \frac{\partial u_{i}}{\partial x_{j}} + \frac{2}{\kappa - 1} a \frac{\partial a}{\partial x_{i}}, \quad 
    u_{j} \frac{\partial a}{\partial x_{j}} + \frac{\kappa - 1}{2} a \frac{\partial u_{j}}{\partial x_{j}}\right)
\end{equation}
\vspace{0.2cm}

A variational formulation is used to create a system of algebraic equations. The residuals of the system are weighted with test functions and integrated over the domain, resulting in high-fidelity system matrices $\bar{\bar{L}}_h$ and $\bar{\bar{Q}}_h$ (see Eq.~\ref{eq:reduced_linear_operator} and Eq.~\ref{eq:reduced_nonlinear_operator}). By reducing the solution space and applying a Galerkin projection as outlined in the introduction, the reduced algebraic systems $\bar{\bar{L}}_r$ and $\bar{\bar{Q}}_r$ can be derived where the reduction of the non-linear parts results in the third-order tensor $Q$. These steps involve time-consuming operations of $O(h)$, but they can be carried out during the offline phase.

\begin{equation}
 	\label{eq:reduced_linear_operator}
 	\bar{\bar{L}} = \bar{\bar{\Phi}}^{T}\bar{\bar{L}}_{h}\bar{\bar{\Phi}}
\end{equation}
\begin{equation}
 	\label{eq:reduced_nonlinear_operator}
 	\bar{\bar{Q}}^{k} = \bar{\bar{\Phi}}^{T}\bar{\bar{Q}}_{h}(\varphi^{k})\bar{\bar{\Phi}}
\end{equation}

If a FE approach is used to derive the high-fidelity systems, the result of this approach is equivalent to directly using the modes for calculating the inner products \cite{Quarteroni.2016}. During the online phase, the non-linear algebraic system is solved with a Newton approach outlined in Eq.~\ref{eq:newton_method}. 

\begin{equation}
    \label{eq:newton_method}
    r^{t+1}_{i} = r^{t}_{i} - \frac{R^{t}_{i}}{\partial R^{t}_{i}/\partial r_{j}^t} = r^{t}_{i} - \frac{L_{ij} r^{t}_{j} +  Q_{ijk} r^{t}_{j} r^{t}_{k}}{L_{ij} +  Q_{ijk} r^{t}_{k}}
\end{equation}

During the execution of the Newton steps, no time-consuming integrations over the computational domain are necessary. As shown in the boundary treatment section, the base $\bar{\bar{\Phi}}=[\bar{\bar{\Omega}}, \bar{\bar{\Psi}}]$ contains both, boundary modes and non boundary modes. The same is true for the vector $\bar{r} = [\bar{b}; \bar{d}]$ where only the entries of $\bar{d}$ are actual DoFs. Thus, during the system evaluation, the columns and rows corresponding to the entries of $\bar{b}$ are removed, and the Newton iterations are only used to update the entries corresponding to $\bar{d}$.

\subsection{Geometrical parametrization}
\begin{figure}[!htpb]
	\centering
	\includegraphics[width=\textwidth]{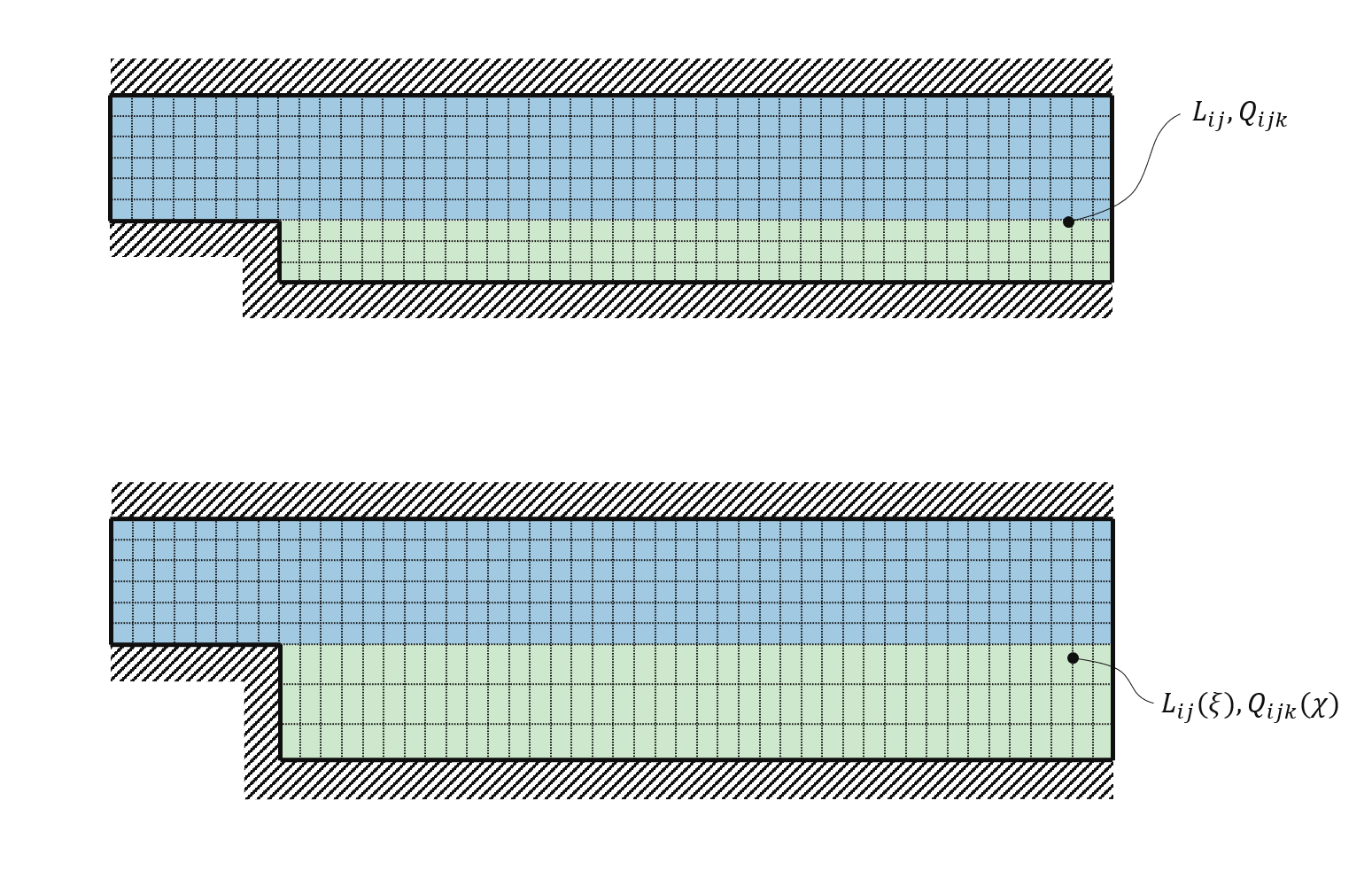}
	\caption{Deformation of computational mesh due to geometric parameter changes. The green elements are stretched and enlarged causing their mapping functions to change.}
	\label{fig:geometrical_parametrization}
\end{figure}
In contrast to physical parameters, where only the scaling factors of the boundary modes need to be adapted (see Eq.~\ref{eq:imposing_boundary_conditions2}), the treatment of geometrical deformations is more challenging. In this case, the computational mesh is deformed (see Figure \ref{fig:geometrical_parametrization}) and thus, node positions and global base function derivatives change. The change of node positions and element sizes leads to changes in the inner products used to assemble tensor $L_r$ and $Q_r$. This means they would need to be reevaluated every time the geometry is modified. This is problematic as it involves time-consuming calculation steps of $O(h)$ and thus undermines the online/offline decoupling of the RB model. 
\\ To circumvent these problems and obtain an efficient RB model, a further layer of hyperreduction is necessary. The approach of this work is similar to the one from Quarteroni und Rozza \cite{Quarteroni.2007} for non-affine parametric deformations. The procedure is based on approximating the tensors $L$ and $Q$ through a linear combination of  $ng$ precalculated tensor bases (see Eq.~\ref{eq:geo_par_linear} and ~\ref{eq:geo_par_nonlinear}).

\begin{equation}
    \label{eq:geo_par_linear}
    L_{ik}(\mu) = \sum^{ng}_{l=1} L_{ikl} \alpha_{l}(\mu)
\end{equation}
\begin{equation}
    \label{eq:geo_par_nonlinear}
    Q_{ijk}(\mu) = \sum^{nq}_{l=1} Q_{ijkl} \beta_{l}(\mu)
\end{equation}

In the first step, the geometry of the sudden expansion channel is defined with a default geometry or reference domain. Every new geometry is a deformation of this default setting and can be mapped forward and back from this reference domain by applying the relations shown in Eq.~\ref{eq:coordinate_transformation}. The partial derivatives are extended using a product rule, and also the integrated is defined on the reference domain and mapped onto the real geometry using the Jacobian determinant.

\begin{equation}
    \label{eq:coordinate_transformation}
    \frac{\partial}{\partial \hat{x}_i} = \frac{\partial x_j}{\partial \hat{x}_i} \frac{\partial}{\partial x_j}, \quad d\Omega = dx \,dy = |\det(\mathbf{J})| d\hat{x} \, d\hat{y}.
\end{equation}

If these relations are inserted into the weak form of the isentropic Navier Stokes system, the arising equations are formulated on the reference domain plus a mapping. The mapping terms $\xi$ and $\chi$ appear next to the linear (see Eq.~\ref{eq:linear_geo_mapping}) and the nonlinear (see Eq.~\ref{eq:quadratic_geo_mapping}) parts of the equations. If the mesh is deformed, the change is reflected by changes in the mapping terms that act linearly onto the equations formulated on the reference domain.

\begin{equation}
    \label{eq:linear_geo_mapping}
    L = L(\xi) = \xi \cdot L_{ref} \quad \rightarrow \quad \xi_{ij} = \frac{\partial x_j}{\partial \hat{x}_i} \frac{\partial x_j}{\partial \hat{x}_i}|\det(\mathbf{J})|
\end{equation}
\begin{equation}
    \label{eq:quadratic_geo_mapping}
    Q = Q(\chi) = \chi \cdot Q_{ref} \quad \rightarrow \quad \chi_{ij} = \frac{\partial x_j}{\partial \hat{x}_i}|\det(\mathbf{J})|
\end{equation}

To simplify the problem, it is assumed that the mapping functions, which are dependent on the coordinates of the reference domain and the current geometrical parametrization, can be expressed as a weighted sum of empirical base functions (see Equations ~\ref{eq:chi_deformation} and ~\ref{eq:xi_deformation}).

\begin{equation}
    \label{eq:chi_deformation}
    \chi(\hat{x}, \mu) = \sum_{m=1}^{ng} \gamma_m(\hat{x}) \alpha_{m}(\mu)
\end{equation}
\begin{equation}
    \label{eq:xi_deformation}
    \xi(\hat{x}, \mu) = \sum_{m=1}^{ng} \zeta_m(\hat{x}) \beta_{m}(\mu)
\end{equation}

If this assumption is introduced into Equations \ref{eq:linear_geo_mapping} and \ref{eq:quadratic_geo_mapping}, this leads to the formulation shown in Eq.~\ref{eq:geo_par_linear} and ~\ref{eq:geo_par_nonlinear}. Parts of the equations can be multiplied out, and as a result, the tensor bases are no longer a function of the geometrical parametrization. This means they can be calculated in advance during the assembly phase, while only $\alpha$ and $\beta$ need to be calculated during the time-critical online phase. For calculations during the online phase, the deformation metrics that led to the different tensor bases are stored in the matrices $\bar{\bar{\chi}}$ and $\bar{\bar{\xi}}$.

\begin{equation}
    \label{eq:linear_geo_mapping2}
    L = \xi \cdot L_{ref} = \left( \sum_{m=1}^{ng} \zeta_m(\hat{x}) \alpha_{m}(\mu) \right) \cdot L_{ref} = \sum_{m=1}^{ng} L_{m} \alpha_{m}(\mu)
\end{equation}
\begin{equation}
    \label{eq:quadratic_geo_mapping2}
    Q = \chi \cdot Q_{ref} = \left( \sum_{m=1}^{ng} \gamma_m(\hat{x}) \beta_{m}(\mu) \right) \cdot Q_{ref} = \sum_{m=1}^{ng} Q_{m} \beta_{m}(\mu)
\end{equation}

To derive $\alpha$ and $\beta$ for a new setup, it is necessary to calculate the deformation metrics $\chi$ and $\xi$ for the current geometry. Finally, a least squares approach projects the current $\chi$ and $\xi$ onto the stored deformation matrices (see Eq.~\ref{eq:calculate_alpha} and Eq.~\ref{eq:calculate_beta}). By doing so, one derives the weighting factors of the bases that lead to the best approximation of the current deformation in terms of the available tensor bases. One drawback of this approach is that it again involves calculations of $O(h)$ even though they are more lightweight. Yet it turns out that in most cases, evaluating the deformation over the full domain is not necessary, but evaluating a small subset of elements is sufficient to accurately "measure" the deformation. In Eq.~\ref{eq:calculate_alpha} and Eq.~\ref{eq:calculate_beta} this is expressed by the sampling matrix $\bar{\bar{P}}_{geo}$ that selects these elements.

\begin{equation}
    \label{eq:calculate_alpha}
    \bar{\alpha}(\mu) = (\bar{\bar{P}}_{geo}^{T} \, \bar{\bar{\xi}})^{+} \, \bar{\bar{P}}_{geo}^{T} \, \xi(\mu)
\end{equation}
\begin{equation}
    \label{eq:calculate_beta}
    \bar{\beta}(\mu) = (\bar{\bar{P}}_{geo}^{T} \, \bar{\bar{\chi}})^{+} \, \bar{\bar{P}}_{geo}^{T} \, \chi(\mu)
\end{equation}

The online operations outlined above do not involve time-consuming steps anymore. Their effort is independent of the mesh size, enabling efficient online/offline decompositions with short evaluation times, even in the case of geometrical deformations.

\section{Results}
\label{sec:Results}
\subsection{RB error convergence (sudden expansion channel)}
\begin{figure}[!htbp]
	\centering
    \vspace{0.7cm}
	\includegraphics[width=\textwidth]{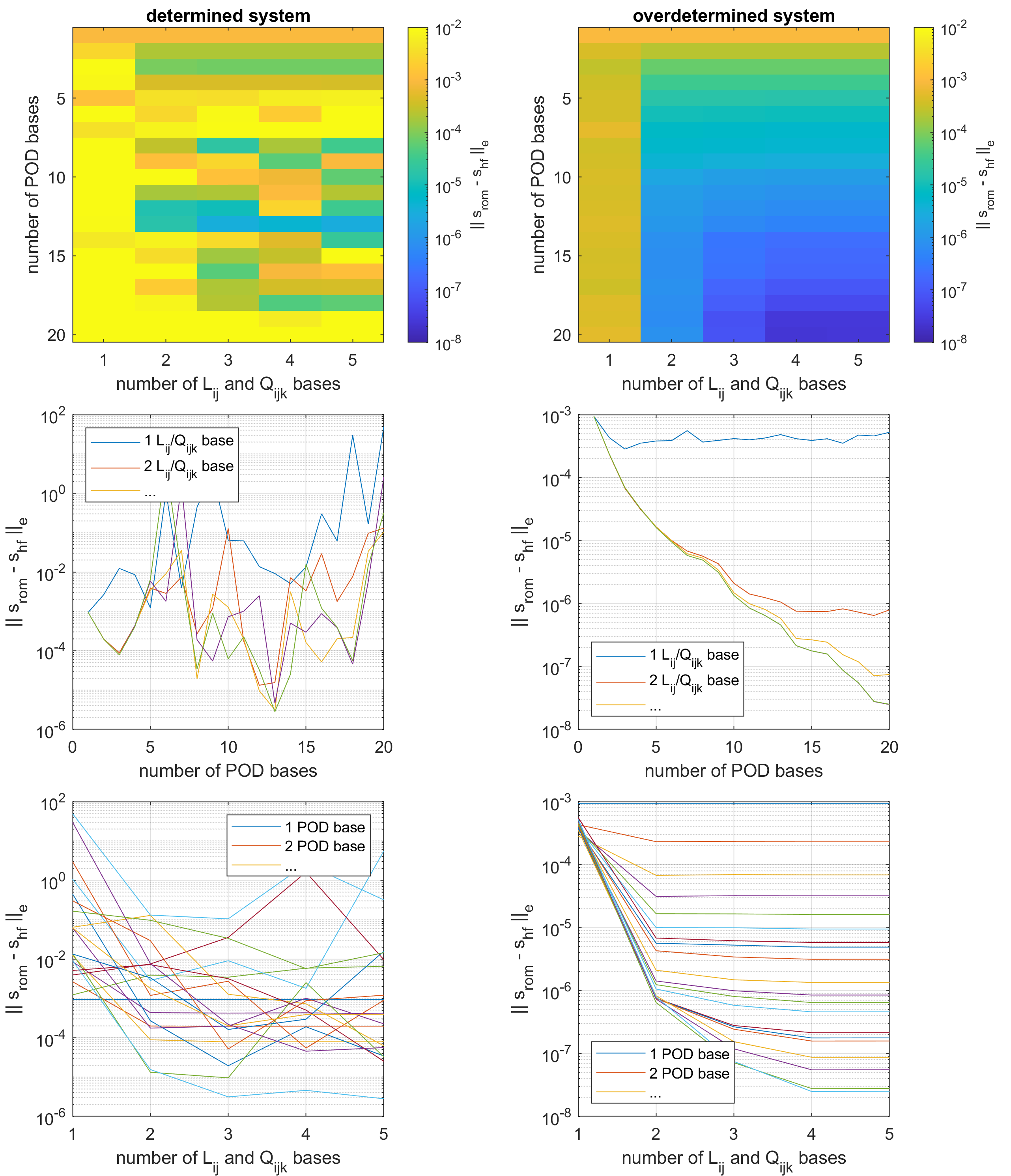}
	\caption{Error convergence for determined (left column) and overdetermined (right column) RB model for different numbers of POD modes and geometrical bases.}
	\label{fig:error_convergence}
\end{figure}
Finally, the implemented RB model is tested on the test dataset of the example case. The test dataset consists of 200 calculations which differ in terms of inlet velocity and step height and thus cover physical and geometrical parameters. The RB model was evaluated for different numbers of geometrical tensor bases and physical modes. The RB model recalculates the 200 test cases with the given physical and geometrical parameter values. At the same time, the RB prediction is compared to the known solution stemming from the stabilized high-fidelity FE model.
\vspace{0.2cm}
\begin{equation}
    e = || \bar{s}_{rom} - \bar{s}_{hf} ||_e
\end{equation}
\vspace{0.2cm}
The energy norm defined in Eq.~\ref{eq:energy_based_prod} is used to evaluate the resulting error vector, giving a scalar value averaged over all 200 test data sets. The test is repeated for a rising number of POD bases and an increasing number of geometrical tensor bases, where the accuracy of the RB model should increase with their numbers: A higher number of orthogonal POD modes increases the dimensionality of the solution space, raising the possibility of expressing variations. When increasing the number of POD modes, the RB prediction should converge to the high-fidelity solution. The corresponding results are given in Figure \ref{fig:error_convergence}. The top row of the figure shows the calculated error values for the different RB models. The columns show how the error behaves if the number of geometrical tensor bases is changed from 1 to 5. The rows show how the error changes if the number of POD modes is increased from 1 to 20. 
\\The solutions in the subplot at the top left corner stem from a standard Galerkin approach with equal trial and test function spaces. This leads to a determined system of equations, and in this case, neither the increase of POD bases nor the increase of geometrical tensor bases increases the accuracy of the result; a higher number of POD bases even worsens the result, leading to diverging solutions and high error amplitudes. 
\\The subplot on the top right corner shows the results for the Petrov-Galerkin model that uses an extended test function space. Instead of using only the truncated POD base, the test function base is extended using truncated modes. This leads to an overdetermined system of equations solved in a least-squares sense in every Newton step. For the given example, the dimensionality of the test function space was double the dimensionality of the trial function space. The RB models show a convergent behaviour if more POD modes are included. 
\\The two subplots in the second row of Figure \ref{fig:error_convergence} show the error for a rising number of POD modes where the different lines are associated with different numbers of geometrical bases. While the determined Galerkin model does not show convergent behaviour, the Petrov-Galerkin model on the right has monotonically decreasing errors for larger solution spaces. The blue line shows an RB model with one geometrical base $L_{ij}/Q_{ijk}=1$. The error of RB models with one tensor base drops slightly if the number of POD modes is increased, but it levels off at high error values. If the number of tensor bases is increased to two, the error drops more than four orders of magnitude before plateauing, while for three or more tensor bases, the plateau has not yet been reached for 20 POD modes. The lines with four and five tensor bases collapse, indicating that four tensor bases are sufficient to reproduce deformations in this simple example case exactly.
\\The third row of Figure \ref{fig:error_convergence} shows the error change for changing tensor base numbers. Again, it can be seen clearly how the left subplot does not show any convergent behaviour, while in the right subplot, the error drops until reaching four tensor bases and levels off afterwards.

\subsection{RB model evaluation times (sudden expansion channel)}
\begin{figure}[!htbp]
	\centering
	\includegraphics[width=\textwidth]{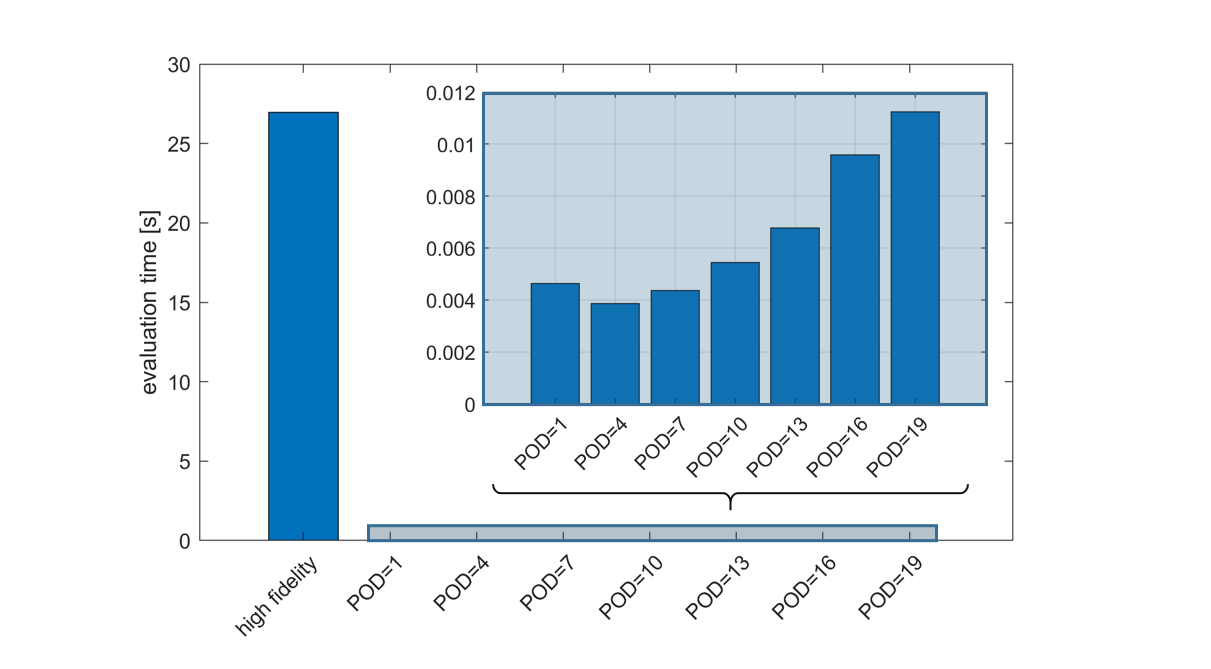}
	\caption{Evaluation times of the high-fidelity model and an enlarged view of multiple RB models with different sizes, ranging from one to 19.}
	\label{fig:evaluation_times}
\end{figure}
One of the main arguments for implementing an RB model is shorter evaluation times. To compare these, the given example case was investigated using the high-fidelity solver and different versions of RB models. For a fair comparison, 15 iterations were performed for each investigated model. The results are given in Figure \ref{fig:evaluation_times}. The high-fidelity model took about 27 seconds for the 15 iterations, while the evaluation times of the different RB models ranged around 0.004 to 0.012 seconds to assemble and solve the system of equations. The models used four tensor bases ($L_{ij}/Q_{ijk}$) and different numbers of POD bases. It can be observed that the evaluation time of the RB models grows progressively with a rising number of POD modes. The MATLAB backslash operator is used to solve the linearized systems in the different iteration steps, which, in the case of a non-square system, uses a QR solver. Depending on the exact procedure, the complexity of this step grows with larger algebraic systems. In the case of a $n \times m$ matrix, the evaluation time for this step is roughly $~O(nm^2)$. The results shown in Figure \ref{fig:evaluation_times} suggest that this might be one of the main drivers for growing computational demand for more accurate models. Other tests, carried out with different numbers of tensor bases and different degrees of overdetermination, did have a smaller influence on the evaluation times. Overall, the speedup of the different investigated models was between three and four orders of magnitude compared to the high-fidelity model. However, one has to keep in mind that all the procedures are implemented in MATLAB, and parts of the code are still open to optimization, which might change results slightly.

\subsection{Adaption to turbulent Flows}
The last adaption of the presented RB model concerns its application to turbulent flows. Up to this point, the model is suited for laminar flow cases with constant viscosity fields. As one goal is to use the model in the context of turbulent flows, the model is slightly modified using a RANS approach. A turbulent viscosity is used to model the effect of unresolved turbulent eddies, changing the viscosity to $\nu_{eff} = \nu_{mol} + \nu_{t}$. This means that the original definition of the linear terms $L$ needs to be modified slightly. It is no longer a constant for unchanged geometries but rather a function of the viscosity field, and thus, the definition of the reduced linear operator changes, as shown in Eq.~\ref{eq:linear_operator_turbulent}.

\begin{equation}
 	\label{eq:linear_operator_turbulent}
 	\bar{\bar{L}} = \bar{\bar{\Phi}}^{T}\bar{\bar{L}}_{h}(\nu_{eff}(x))\bar{\bar{\Phi}}
\end{equation}

Again, it is assumed that the viscosity field can be decomposed into viscous modes denoted by $\pi$ (see Eq.~\ref{eq:viscos_modes}) where the procedure is analogue to the one shown in Eq.~\ref{eq:continuous_solution_space}. Next, the assumption is introduced into the definition of the reduced linear operator. This adds a third index direction to the initial 2-dimensional linear operator.

\begin{equation}
    \nu_{eff}(x) = \sum^{k}_{i=1} \pi_k(x) \, p_k(\mu) \quad \rightarrow \quad \bar{\nu}_{eff}=\bar{\bar{\Pi}} \, \bar{p}
    \label{eq:viscos_modes}
\end{equation}
\begin{equation}
     \bar{\bar{L}}_{k} = \bar{\bar{\Phi}}^{T}\bar{\bar{L}}_{h}(\bar{\pi}_{k})\bar{\bar{\Phi}}
\end{equation}

During the Newton iterations, this operator is contracted along this index direction using the turbulent viscosity field's reduced representation $\bar{p}$. This changes the initial formulation of each Newton step shown in Eq.~\ref{eq:newton_method} to the formulation below.

\begin{equation}
    \label{eq:newton_method_turbulent}
    r^{t+1}_{i} = r^{t}_{i} - \frac{R^{t}_{i}}{\partial R^{t}_{i}/\partial r_{j}^t} = r^{t}_{i} - \frac{L_{ij} r^{t}_{j} +  Q_{ijk} r^{t}_{j} r^{t}_{k}}{L_{ijk} p_{k}+  Q_{ijk} r^{t}_{k}}
\end{equation}

There are multiple options to determine the entries of $\bar{v}$, including the evaluation of conservation equations, but for the presented example, a data-driven approach was chosen. The coefficients for the turbulence field are derived by training a feed-forward neural network that takes the inlet parameters as input and returns the vector entries, creating a mapping $\bar{v} = n(\bar{\mu})$ where $n$ is the neural network. The network uses a simple feed-forward architecture with tanh activation functions and three hidden layers with 20, 15 and 20 neurons, respectively. It was trained with 200 data sets using a Levenberg-Marquard algorithm. Due to its small size, the network's training only took a few seconds. Finally, the network was tested on the remaining 100 data sets with good agreement.

\subsection{Turbulent flow setup}
\begin{figure}
	\centering
	\includegraphics[width=\textwidth]{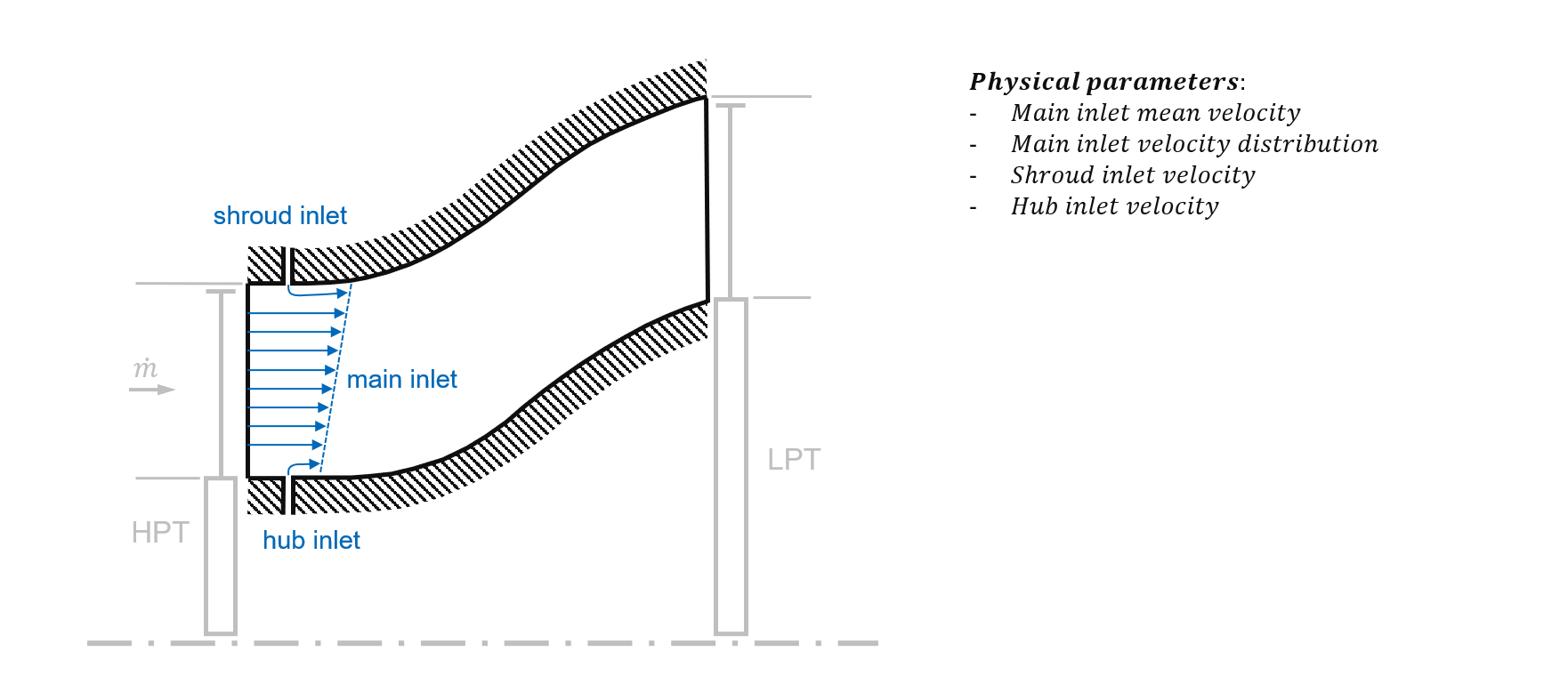}
	\caption{Setup for turbulent flow simulation through a simplified S-shaped duct with three inlets through the channel and the hub and shroud cavities. The outlet boundaries are unchanged.}
	\label{fig:turbulent_flow_setup}
\end{figure}

The turbulent flow simulations are tested on a two-dimensional simplification of a turbine center frame which is the flow channel between the high- and low-pressure turbine (HPT/LPT) in high-bypass aero engines (see Figure \ref{fig:turbulent_flow_setup}). The duct is S-shaped, routing the flow from the low HPT radius to the higher LPT radius. The flow enters the geometry through the main inlet and two further inlets at the hub and shroud of the duct. Both represent simplified versions of HPT cavities, which are pressurized with sealing air from the compressor to prevent hot gas ingress. As the sealing between the turbine and the channel is not entirely tight, this purge air enters the main channel through the cavities and mixes with the main flow. 
The flow setup is defined in a four-dimensional parameter space. Two parameters specify the mean inlet velocity through the main channel and the inclination of this velocity profile. The inclination parameter is defined such that a value of zero means a constant velocity profile, where a value of one means that the velocity at the shroud is twice the mean velocity while the velocity at the hub is zero. The remaining two parameters are the hub and shroud inlet velocity, respectively. The parameter range is given in the table below and chosen so that the highest velocities within the domain are around 120 [m/s].
\vspace{0.5cm}
\begin{center}
\captionof{table}{S-duct parameter range}
   \begin{tabular}{|c|c|c|}
   \hline
    Parameter name & Min. value & Max. value \\
    \hline
    Main inlet mean & 50 [m/s] & 90 [m/s] \\
    Main inlet inclination & 0 [1] & 0.5 [1] \\
    Hub inlet & 0 [m/s] & 60 [m/s] \\
    Shroud inlet & 0 [m/s] & 60 [m/s] \\
    \hline
    \end{tabular} 
\end{center}
\vspace{0.5cm}
Again, the sampling of the parameter space is carried out using a random approach, and the CFD calculations are carried out using ANSYS Fluent. The simulations are carried out without an energy equation, and the density is defined by the isentropic relations shown in Eq.~\ref{eq:isentropic_relations}. A structured mesh is created using the ICEM meshing tool, and the cell size is set such that the smallest cell reaches a y+ value of around one. The database created with ANSYS contains 300 simulations, and 100 simulations are used to evaluate the accuracy of the RB model.

\subsection{Turbulent flow results}
\begin{figure}[!b]
	\centering
	\includegraphics[width=\textwidth]{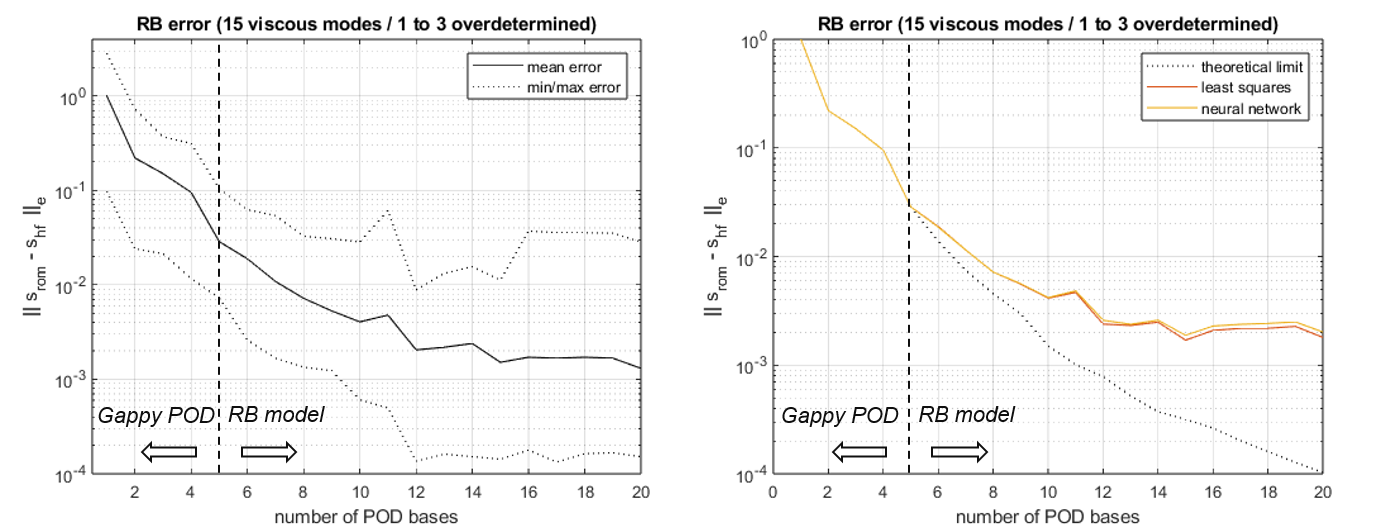}
	\caption{Error convergence for turbulent flow setup through an S-shaped duct with mean/min/max values (left) and in comparison with the theoretical limit (right).}
	\label{fig:turbulent_flow_error_convergence_mean_max}
\end{figure}

\begin{figure}
	\centering
	\includegraphics[width=\textwidth]{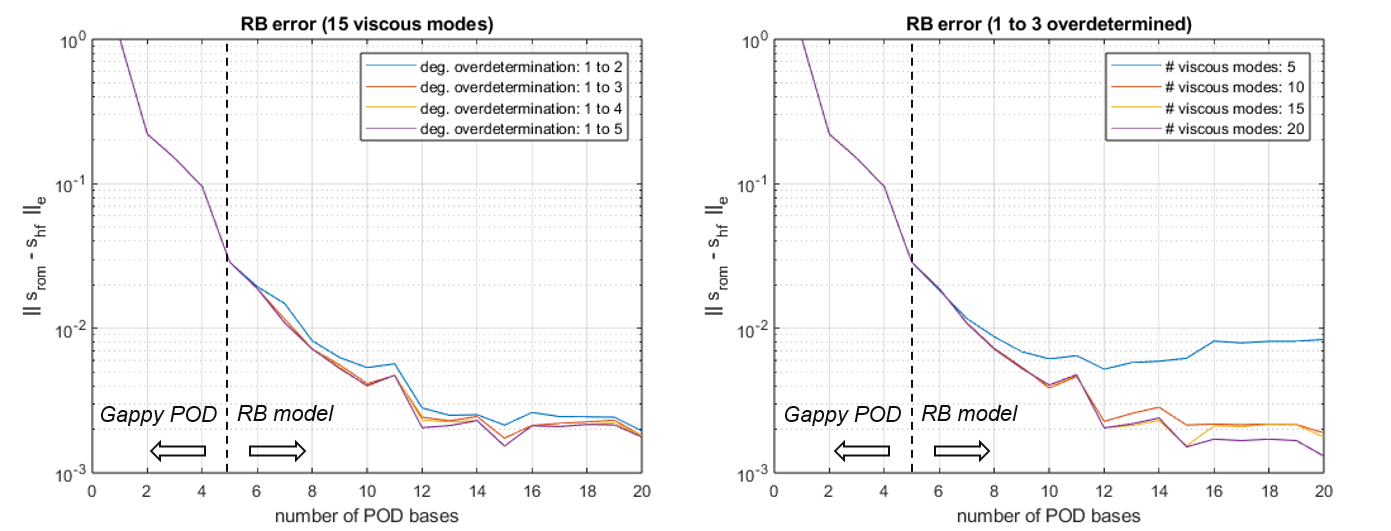}
	\caption{Error convergence for turbulent flow setup through s-shaped duct, showing the trend for different degrees of over determination (left) and different numbers of viscous modes (right).}
	\label{fig:turbulent_flow_error_convergence_overdet}
\end{figure}

The convergence analysis results are shown in Figures \ref{fig:turbulent_flow_error_convergence_mean_max} and \ref{fig:turbulent_flow_error_convergence_overdet}. All plots show the decay of a scalar error measure with a growing number of POD modes. The error measure is calculated by subtracting the RB model prediction from the Fluent solution in each mesh node. Then, the differences are weighted according to the element sizes, following the approach proposed in Eq.~\ref{eq:energy_based_prod}. This gives an integral error value for each result. The error measure is calculated and averaged over the 100 test cases and normalized by the first value so that the largest deviation is one. In Figure \ref{fig:turbulent_flow_error_convergence_mean_max} left, the mean, maximum and minimum errors are shown for a model with 15 viscous modes (see Eq.~\ref{eq:viscos_modes}) and a degree of overdetermination of three (the number of test functions is three times the number of trial functions). The dotted vertical line marks the location from where the RB model solves the governing equations. All models with less than five POD modes predict the flow field based on a gappy POD approach applied to the boundary conditions. In comparison, RB models with more than five POD modes use the boundary conditions to fit the first five modes and solve the governing equations for the remaining modes.
\\The errors initially drop fast, but the decay levels after reaching 12 modes. The RB model seems to converge to slightly different results, while a qualitative investigation will be shown later. The subplot on the right of Figure \ref{fig:turbulent_flow_error_convergence_mean_max} shows the performance of the RB model in comparison with the theoretical limit of the chosen POD base: Due to the reduced solution space, even if the perfect scaling factors $r_i$ were found, the reconstructed flow field wouldn't be precisely the same as the one from the high-fidelity solution. The black dotted line shows the error of this best possible solution and its continuous rate of decrease while the RB model predictions level off at higher error rates. The remaining two lines show the error introduced by the reconstruction of the turbulent field due to the interpolation with a neural network. The "least squares" line shows the error when the viscous field of the test case is projected onto the available viscous modes (in this case, 15 modes) in a least-squares sense and thus giving the best possible $\bar{p}$ values for each case. The "neural network" line shows the performance of the RB model when the viscous scaling factors $\bar{p}$ (see Eq.~\ref{eq:viscos_modes}) are interpolated using a neural network. Both methods show very similar trends.
In Figure \ref{fig:turbulent_flow_error_convergence_overdet} left, the RB accuracy is investigated for different degrees of over-determination. Here, the case for a determined system is missing as it did not converge and produced very high error measures above 100. The curves show that increasing the size of the test function space is beneficial, but the strength of the effect is negligible for high degrees of over-determination above two. In the right subplot, the RB model accuracy is tested for different numbers of viscous modes. Again, a clear trend can be observed where a higher number of viscous modes leads to a better agreement between the RB prediction and the Fluent result. A large drop can be observed if the number is raised from five to 10 while a further increase to 15 and 20 only shows a small improvement. The trend seems reasonable as more viscous modes allow for a more accurate reconstruction of the "true" viscosity field.
\begin{figure}[h]
	\centering
	\includegraphics[width=\textwidth]{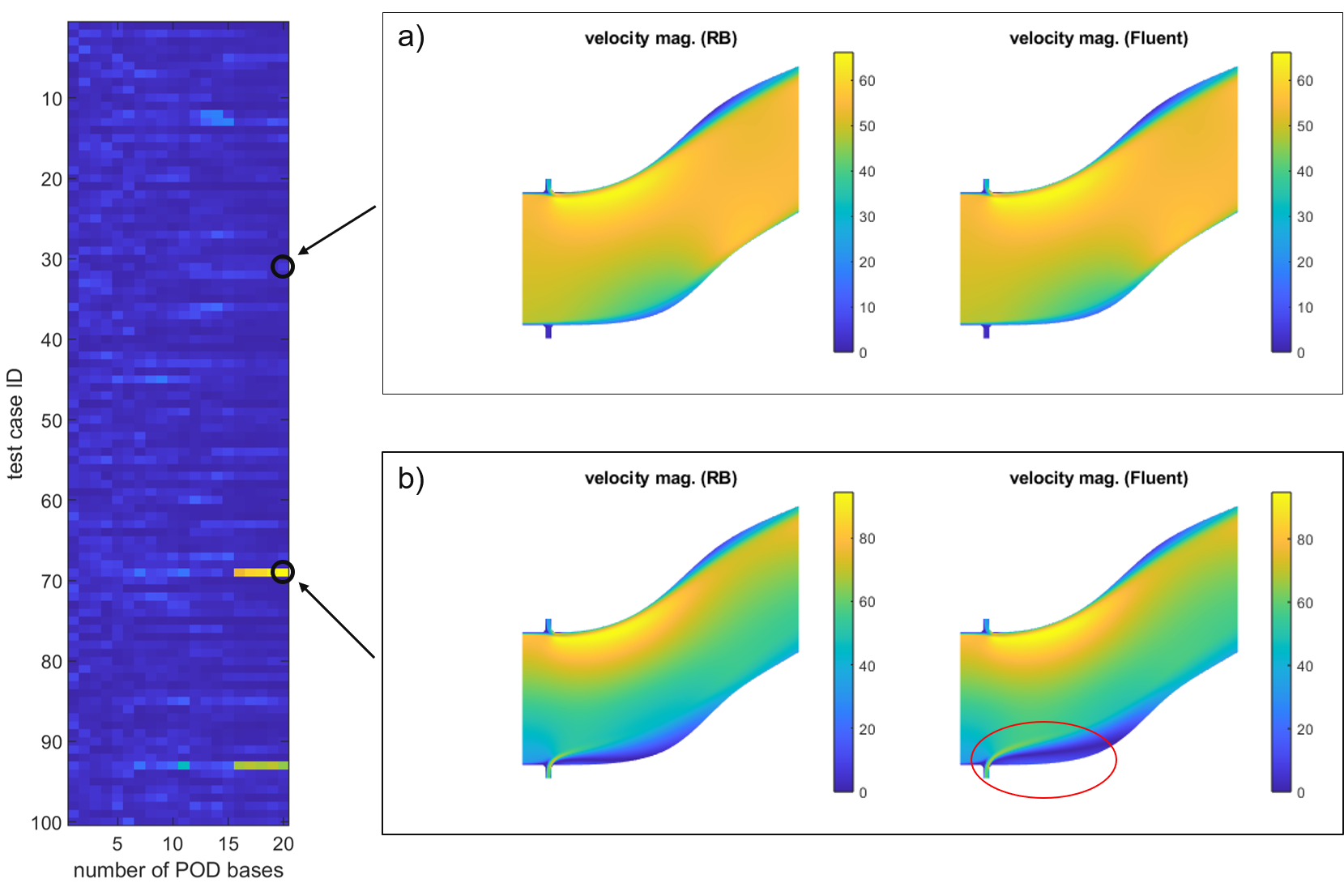}
	\caption{Error values for each test case (left) and Flow field comparisons for the worst case (a) and a further representative case (b).}
	\label{fig:turbulent_flow_error_qualitative}
\end{figure}
\\Figure \ref{fig:turbulent_flow_error_qualitative} shows a qualitative investigation of the 100 test cases. The left subplot shows the error measure of each test case over the individual test case ID and the number of POD modes. The colour indicates the size of the error of each test case, where test cases coloured in blue have small errors while test cases coloured in yellow have larger deviations. Each column is normalized, and the goal of this plot is to show that for the most accurate models with 15 or more POD modes, the remaining errors stem mainly from a few cases that exhibit larger deviations from the Fluent solution. The subplots on the right side show the velocity fields for two examples where subplot a) is chosen to show a representative case with an average error value while subplot b) shows the case with the highest deviation. For case a), the agreement is generally good, and it is difficult to spot differences with the bare eye. The agreement between RB prediction and Fluent result is equally good for 98 out of 100 test cases. For two cases, however, larger deviations were observed, and the velocity field for the worst case is shown in subplot b). The biggest differences are observable near the hub surface, where the size of the recirculation zone, predicted by the RB model, is smaller than the size predicted by the Fluent simulation. In this sensitive case, both methods converge to significantly different results. Both "high-error" cases are located on the edge of the investigated parameter space.

\section{Discussion and conclusion}
\label{sec:Discussion}
In the course of the work, the implementation of an RB model suited for compressible flow was presented. The model allows a parametrization of boundary conditions and the geometry, and it is tested on two different cases. In the first case, the model is applied to the canonical example of a sudden expansion channel where the parameter space includes the inlet velocity and the height of the step. To assess the reliability of the model, it was investigated concerning the deviation from a high-fidelity CFD simulation for different numbers of POD modes and geometrical tensor bases; in theory, a higher number of bases should improve the accuracy of the model as the solution space allows a better approximation of the actual result. But in the case of a standard Galerkin model, no clear trend could be observed, and some solutions diverged and, thus, even worsened the accuracy of the RB model for a growing number of POD bases. However, if a Petrov-Galerking approach with an enriched test space is used, the model's predictions are reliable and follow the expected trend: when reaching 20 POD modes, the error of the RB model has dropped by almost five orders of magnitude, and results are indistinguishable with the bare eye. In this case, the stabilization was achieved using trial and test spaces of different dimensionality. This leads to an overdetermined system of equations in each Newton step, yet due to the very small size of the algebraic systems, the increased effort to solve these dense, overdetermined systems is still small. For all investigated models, the calculation times were multiple orders of magnitude lower than for the high-fidelity calculation. The results of this example and the error drop rate are comparable to other works from the literature, where the same setup was investigated for incompressible flow. 
\\In the second example, the implemented model is applied to the turbulent flow through a simplified two-dimensional S-shaped duct, as found in high-bypass aero engines. The geometry was unchanged in this case, but the boundaries are described with four different parameters. The RB model is extended to accommodate turbulent flows in the RANS framework by allowing a non-constant effective viscosity over the domain. In this case, the reference CFD calculations are not generated by the same CFD-solver as the RB model but by using the commercial software ANSYS Fluent. Again, the RB model is evaluated in terms of accuracy by varying the number of POD modes, the degree of overdetermination and the number of viscous modes. Also, in this case, the deviation between the high-fidelity case and the RB prediction drops if the number of POD modes is increased, but it plateaus after being dropped about three orders of magnitude. For the investigated case, the number of five viscous modes was insufficient to generate accurate results, and a big increase in accuracy could be observed if the number of viscous modes was raised from five to 10. For even higher values, the trend persisted but weakened. A higher degree of overdetermination also had a beneficial effect but was less pronounced than the effect of the viscous field. Other possible reasons for the remaining deviations have been investigated, even though they are not included in the results as their influence appeared to be relatively small:
\begin{itemize}
    \item The mesh independence of the remaining error was investigated by repeating the investigations using a finer mesh for the Fluent simulations and the RB model. The finer mesh was found to have a weak beneficial effect.
    \item The influence of the isentropic assumption for the CFD calculations was investigated. The Fluent calculations were carried out using an ideal gas, isentropic, and incompressible equations, while the RB model always relied on the isentropic equations. As different governing equations are used for the Fluent and RB simulations, this leads to inconsistencies. Yet, only small differences could be observed if an ideal gas was used for the Fluent simulations. While larger differences were found for the incompressible case.
    \item Another big difference is that the RB model is based on an FE discretization method, while Fluent uses an FV formulation and interpolates the values at the nodes. Furthermore, it cannot be said with certainty that the Fluent simulations represent the "true" solution: Even if two high-fidelity CFD codes are compared, it is to be expected that there are differences between their predictions, especially in strongly non-linear flow regimes.
\end{itemize}
The investigations suggest that there is no single explanation for the remaining deviations but that all of them would need to be addressed simultaneously to increase the RB model's accuracy further. Still, it has to be mentioned here that the deviations are relatively small and that the few cases where more significant deviations could be observed did not occur randomly but are consistent when looking at the different RB models with varying POD base numbers.
\\For future applications, the RB model needs to be extended to three dimensions and also the currently implemented version to determine the viscosity fields is still open to improvements. Formulating it without using regression techniques could be advantageous when it comes to extrapolations and a possibility to achieve this is to embed a turbulence model into the RB framework, using a DEIM approach for a fast estimation. Overall, short computation times are a big advantage of RB models. If the model is designed for multi-query applications, the initial costs during the costly offline phase are marginalized over time. Furthermore, the compact RB models allow traceable calculations with so-called dlarrays, enabling fast and exact automatic differentiation to calculate gradients for optimization tasks.

\section*{Acknowledgments}
The authors would like to acknowledge the financial support of the "ARIADNE" project by the "Take Off" Program, a Research, Technology, and Innovation Funding Program of the Republic of Austria's Ministry of Climate Action. The "ARIADNE" Programme is managed by the Austrian Research Promotion Agency (FFG).

\bibliographystyle{ieeetr}
  \bibliography{main.bib}

\end{document}